\documentclass[sigconf]{acmart}

\AtBeginDocument{%
  }

\setcopyright{acmlicensed}
\copyrightyear{2026}
\acmYear{2026}
\setcopyright{cc}
\setcctype{by}
\acmConference[HAI '26]{Proceedings of the 14th International
Conference on Human-Agent Interaction}{November 16--19, 2026}{Osaka,
Japan}
\acmBooktitle{Proceedings of the 14th International Conference on
Human-Agent Interaction (HAI '26), November 16--19, 2026, Osaka,
Japan}
\acmDOI{10.1145/3841580.3841626}
\acmISBN{979-8-4007-2575-3/2026/11}

\newcommand{\CCV}{\mathrm{CCV}}
\newcommand{\Ldir}{L_T^{\mathrm{direct}}}
\newcommand{\Lfb}{L_T^{\mathrm{feedback}}}
\newcommand{\LMone}{L_T^{\mathrm{M1}}}
\newcommand{\LMtwo}{L_T^{\mathrm{M2}}}
\newcommand{\Sgeo}{S^{\mathrm{GEO}}}

\begin{document}

\title[Conversational Capture: Trajectory-Level GEO Evaluation]{%
Conversational Capture: A Trajectory-Level Framework for Evaluating Generative Engine Optimization in Multi-turn Human-Agent Interaction}

\author{Junwei Yu}
\orcid{0009-0004-1657-3310}
\affiliation{%
  \institution{The University of Tokyo}
  \city{Tokyo}
  \country{Japan}
}
\email{yujw@satolab.itc.u-tokyo.ac.jp}

\author{Jieyu Zhou}
\orcid{0009-0005-9346-5214}
\affiliation{%
  \institution{UniConvo Inc.}
  \city{Tokyo}
  \country{Japan}
}
\email{zhou@uniconvo.co.jp}

\author{Mufeng Yang}
\orcid{0009-0005-1714-2544}
\affiliation{%
  \institution{University of Tsukuba}
  \city{Tsukuba}
  \state{Ibaraki}
  \country{Japan}
}
\email{s2321728@u.tsukuba.ac.jp}

\author{Yepeng Ding}
\correspondingauthor
\orcid{0000-0002-6996-9333}
\affiliation{%
  \institution{Hiroshima University}
  \city{Higashihiroshima}
  \country{Japan}
}
\email{ypding@hiroshima-u.ac.jp}

\author{Hiroyuki Sato}
\orcid{0000-0002-2891-3835}
\affiliation{%
  \institution{National Institute of Informatics}
  \city{Tokyo}
  \country{Japan}
}
\email{schuko@nii.ac.jp}

\renewcommand{\shortauthors}{Yu et al.}

\begin{abstract}
Generative Engine Optimization (GEO) shapes content to increase its likelihood
of being cited by answer engines built on retrieval-augmented large language
models. GEO is typically evaluated as a single-turn property: for a fixed
query, an evaluator measures a source's visibility in one answer. We argue that
the single answer is an inadequate unit of analysis. Human-agent information
seeking forms a closed loop. The agent's answer changes the user's beliefs and
therefore the next question, which in turn determines what the agent retrieves.
We introduce conversational capture, a phenomenon in which a source cited
early becomes substantially more likely to be cited again. Capture operates
through a machine-side channel, history-conditioned retrieval, and a human-side
channel, follow-up questions directed toward the captured source. This
persistence can be independent of the source's current relevance. We formalize
the interaction as a two-layer closed-loop system and derive computable
trajectory-level constructs: cumulative conversational visibility; a
decomposition of trajectory gain into a single-turn "direct" term and a
closed-loop "feedback" term; a nested counterfactual split of the feedback
term into machine-side and human-side channels; a capture coefficient;
a compounding ratio; and a misranking diagnostic. Using reinforcement-process
(Pólya-urn) theory, we prove that the feedback term is identically zero under
single-turn evaluation and that GEO's cumulative payoff grows superlinearly
with conversation length while capture develops. A model-derived illustration
shows that the feedback term can exceed the direct term, the compounding ratio
exceeds two within ten turns, and the single-turn and trajectory rankings of
GEO methods agree only weakly (Kendall's $\tau = 0.4$). Thus, single-turn
evaluation can select the wrong method. We connect the human channel to
information foraging, trust calibration, and Bayesian persuasion, and discuss
design implications for answer engines.
\end{abstract}



\begin{teaserfigure}
  \centering
  \includegraphics[width=\textwidth]{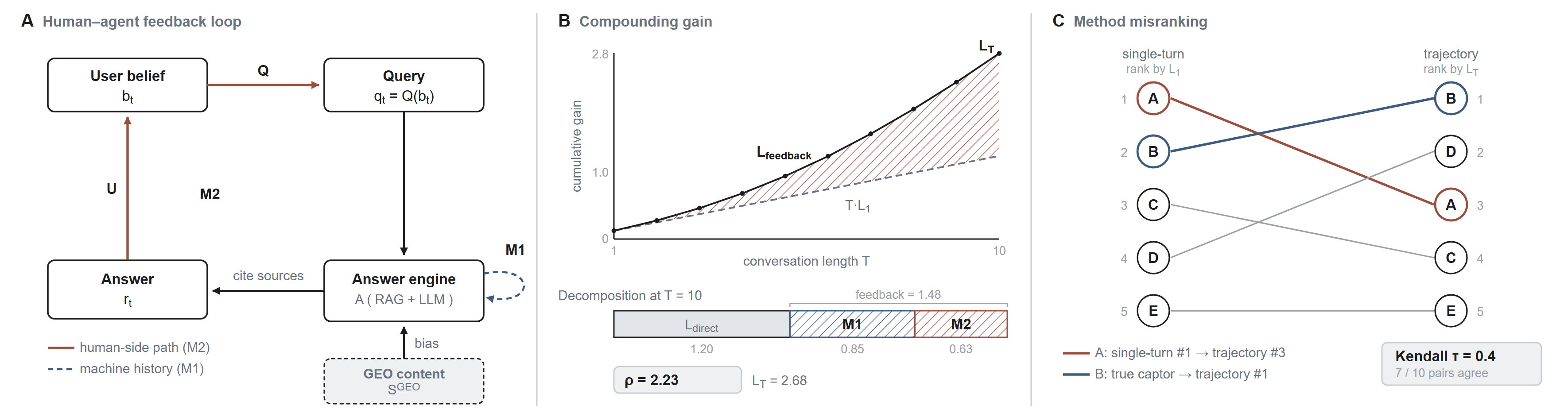}
  \caption{Overview of conversational capture. \textbf{(A)} Human-agent
  information seeking is a closed loop: the answer $r_t$ updates the user's
  belief $b_{t+1}$ and hence the next query $q_t=Q(b_t)$, where $Q$ is the
  query operator (human-side channel
  \textsf{M2}), while also conditioning the engine on the dialogue history
  $h_t$ (machine-side channel \textsf{M1}); an initial GEO bias on the optimized
  source set $\Sgeo$
  propagates through both return paths. \textbf{(B)} Because the loop
  compounds, cumulative trajectory gain $L_T$ exceeds the single-turn
  extrapolation $T\cdot L_1$; the excess is the feedback term $\Lfb$, which is
  identically zero under single-turn evaluation. The bar splits $L_T=2.68$ at
  $T{=}10$ into the direct term ($\Ldir=1.20$) and the feedback term
  ($\Lfb=1.48$). The latter is divided into its machine-side channel
  ($\LMone=0.85$, blue, no user
  model required) and human-side channel ($\LMtwo=0.63$, rust), coloured as in
  (A); the compounding ratio is $\rho=2.23$. \textbf{(C)} Consequently, ranking
  GEO methods by single-turn gain disagrees with ranking them by trajectory
  gain (Kendall's $\tau = 0.4$): the single-turn winner (A) is only third on
  trajectory.}
  \Description{A three-panel overview figure. Panel A shows a directed cycle
  of user belief, query, answer engine, and answer, with a human-side return
  path (rust, M2) and a machine-side history self-loop (blue, M1), plus a GEO
  content node biasing the engine. Panel B shows a chart in which the
  trajectory-gain curve bows above a straight single-turn extrapolation line,
  with the gap labeled as the feedback term, and below it a horizontal stacked
  bar decomposing the trajectory gain into three segments: a grey direct term
  of 1.20, a blue machine-side feedback segment (M1) of 0.85, and a rust
  human-side feedback segment (M2) of 0.63, the latter two spanned by a bracket
  labeled feedback equals 1.48, with the compounding ratio rho equals 2.23 and
  total gain 2.68 shown alongside. Panel C shows two ranked columns of five GEO
  methods with crossing connector lines. The crossings indicate that single-turn and
  trajectory rankings disagree, annotated with Kendall tau equals 0.4.}
  \label{fig:overview}
\end{teaserfigure}

\maketitle

\section{Introduction}

People increasingly obtain information by conversing with an \emph{answer
engine} rather than by selecting links from a ranked list. An answer engine
combines a large language model (LLM) with a retrieval component to produce
natural-language responses that often cite sources. This shift has prompted
\emph{Generative Engine Optimization} (GEO), the practice of adapting content
so that answer engines surface and cite it preferentially~\cite{aggarwal2024geo}.
As with search engine optimization, visibility in a generated answer attracts
attention and therefore has value.

Existing work measures GEO's impact in a single-turn setting: it fixes a query,
runs the engine, and scores the prominence of a target source in that answer
using position-adjusted word count, an LLM-judged impression score, or a related
metric~\cite{aggarwal2024geo,kumar2024manipulating}. This approach assumes that
the query is \emph{exogenous}, meaning that it is supplied independently of the
system, and that turns are \emph{independent}. A method's value is therefore
represented by its average lift per query.

These assumptions do not hold in a conversation. Human-agent information
seeking forms a \emph{closed loop}: the user asks a question, reads the answer,
and adjusts the next question in response. The agent's reply thus influences
rather than merely answers the next query. A GEO bias toward a source can
propagate by steering the user's follow-up toward that source and by
conditioning subsequent retrieval on a history that already privileges it.
The resulting advantage can compound over turns.

\paragraph{Thesis.} \emph{Single-turn visibility metrics systematically
misestimate the true impact of GEO because they ignore the human-agent
feedback loop intrinsic to multi-turn dialogue. The real return to GEO is not
single-turn exposure, but its cumulative, path-dependent influence on the
user's sequence of interactions.}

We name the central phenomenon \emph{conversational capture}: in multi-turn
generative retrieval, a source cited in earlier turns has a substantially higher
probability of being cited in later ones, and this persistence can be
\emph{decoupled from its relevance} to the user's current need, which may have
narrowed or shifted. Capture has two distinct causes that we keep separate
throughout the paper. On the \emph{machine side}, retrieval and generation
condition on the dialogue history, so a previously cited source can resurface
regardless of the user~\cite{liu2024lostmiddle}. On the \emph{human side}, a
GEO-shaped answer directs the user's next question toward the captured source.
This channel exists only when a human participates in the loop. Machine-side
capture is a measurable system property, whereas human-side capture raises a
human-computer interaction problem.

\paragraph{Gap in existing work.} The original GEO formulation represents its
input as a dialogue history and therefore permits a multi-turn setting.
However, it does not examine the consequences of that setting for evaluation.
A history that is endogenously produced by the system's previous outputs differs
from a fixed input because the treatment can change later queries. We address
this gap by showing how such endogeneity invalidates the prevailing evaluation
paradigm under multi-turn conditions and by providing a trajectory-level
alternative. We do not propose a new GEO method or benchmark.
Figure~\ref{fig:overview} summarizes the argument.

\paragraph{Contributions.}
\begin{itemize}
  \item \textbf{A concept (C1).} We introduce \emph{conversational capture}
  and reframe GEO as a path-dependence phenomenon in a multi-turn human-agent
  feedback loop, rather than a single-turn exposure phenomenon
  (Section~\ref{sec:motivation}).
  \item \textbf{A model (C2).} We formalize the interaction as a two-layer
  closed-loop dynamical system and separate two independent capture
  channels, machine-side (M1) and human-side (M2), and relate the human
  channel to information foraging, trust calibration, and Bayesian persuasion
  (Sections~\ref{sec:model} and~\ref{sec:hci}).
  \item \textbf{An evaluation framework (C3).} We define a family of
  computable, trajectory-level constructs: cumulative conversational
  visibility, a direct/feedback decomposition of trajectory gain whose feedback
  term is split by a nested counterfactual into its machine-side (M1) and
  human-side (M2) channels, a capture
  coefficient, a compounding ratio, and a misranking diagnostic
  (Section~\ref{sec:constructs}).
  \item \textbf{A theory (C4).} Using reinforcement-process (Pólya-urn)
  theory, we prove that the feedback term vanishes under single-turn
  evaluation, that capture induces non-ergodic path dependence, and that
  cumulative GEO gain grows superlinearly in conversation length while capture
  develops. Single-turn extrapolation underestimates this gain by a factor that rises
  toward a finite ceiling (Section~\ref{sec:theory}).
  \item \textbf{A demonstration (C5).} A model-derived computational
  illustration shows that every construct is computable and that single-turn
  and trajectory-level rankings of GEO methods can diverge (Kendall's
  $\tau = 0.4$). This is a property of the model rather than a measurement of a
  deployed engine. It establishes that the assumptions of the model do not
  imply the sufficiency of single-turn evaluation
  (Section~\ref{sec:illustration}).
\end{itemize}

This paper contributes theory and modeling rather than a new human study. We
collect no human annotations. The computational illustration is a controlled,
reproducible calculation from the model's dynamics. Section~\ref{sec:limitations}
identifies the claims that require empirical validation with deployed retrieval
systems and real users.

\section{Motivation: Why Single-Turn GEO Evaluation Fails}
\label{sec:motivation}

\subsection{The implicit assumptions of current evaluation}
The GEO literature operationalizes ``impact'' as a single-turn visibility
lift~\cite{aggarwal2024geo,kumar2024manipulating}. The standard procedure
samples a query $q$ from a distribution, runs the engine with and without the
optimized content, and reports the difference in a visibility score, such as
position-adjusted word count or an LLM-judged impression. Averaging over
independently sampled queries yields the reported estimate. Two assumptions
support this procedure: queries
are \emph{exogenous} (drawn from a fixed distribution unaffected by the
engine), and turns are \emph{independent} (each query is scored in isolation).

\subsection{The multi-turn reality: a closed loop}
Neither assumption holds in a conversation. Let $b_t$ denote the user's belief
or information state at turn $t$, $q_t$ the query, and $r_t$ the answer. The
interaction proceeds as
\[
b_t \;\xrightarrow{\;\text{ask}\;}\; q_t
\;\xrightarrow{\;\text{answer}\;}\; r_t
\;\xrightarrow{\;\text{update}\;}\; b_{t+1}
\;\xrightarrow{\;\text{ask}\;}\; q_{t+1} \;\cdots
\]
The user's next question is shaped by the agent's last answer. If GEO has
biased $r_t$ toward a source $s$, that bias does two things. It conditions the
engine's future retrieval on a history in which $s$ is already prominent, and
it nudges $b_{t+1}$, and hence $q_{t+1}$, toward the region of information
space that $s$ occupies. Both effects raise the chance that $s$ is cited at
turn $t+1$, which feeds the same loop again. Queries are \emph{endogenous}, and
turns are \emph{coupled}.

\subsection{Conversational capture}
We summarize the consequence as a named phenomenon.

\begin{quote}
\emph{Conversational capture.} In multi-turn generative retrieval, once a
source is cited in earlier turns, its probability of being cited in later
turns is significantly elevated; this persistence may be independent of the
source's relevance to the user's current (narrowed or drifted) information
need.
\end{quote}

\noindent Three mechanisms produce it: (i) the dialogue history is part of the
model's conditioning context; (ii) models tend to remain consistent with their
own prior answers~\cite{sharma2023sycophancy}; and (iii) retrieval itself is
biased by conversational context. The implication for evaluation is direct:
if ``GEO return $=$ single-turn exposure,'' then capture is
invisible. A source that gains an early advantage can retain that advantage
throughout the conversation. This is a claim about trajectories rather than
individual turns. It connects GEO to long-standing concerns about search-engine
bias and filter bubbles~\cite{introna2000shaping,pariser2011filter,epstein2015search,sharma2024generative},
but locates the mechanism in the turn-to-turn dynamics of generative dialogue.

\section{Positioning and Related Work}
\label{sec:related}

\paragraph{GEO and adversarial visibility.} GEO was introduced as a
black-box method for optimizing source content for generative
engines~\cite{aggarwal2024geo}; related work shows content can be manipulated
to inflate product visibility in LLM answers~\cite{kumar2024manipulating}.
This work measures visibility per answer. We instead examine the \emph{unit of
evaluation}. We show that the single turn is misspecified for conversational
settings and propose a trajectory-level alternative.

\paragraph{Conversational search and retrieval-augmented generation.}
Open-retrieval conversational
question answering~\cite{qu2020orconvqa,anantha2021qrecc,adlakha2022topiocqa}
and retrieval-augmented generation (RAG)~\cite{lewis2020rag,karpukhin2020dpr}
establish that retrieval is conditioned on dialogue context and that LLMs
attend unevenly to that context~\cite{liu2024lostmiddle}. These findings provide
the basis for the machine-side capture channel. Multi-turn evaluation is
not new in this sense: these benchmarks already score systems across complete
dialogues with a growing history. However, they replay a \emph{fixed},
prerecorded query sequence, so the path remains exogenous and the treatment
cannot change subsequent questions. In such a replay, $\LMtwo$ is zero by
construction, while $\LMone$ is present but not separated from the direct term.
Our claim therefore concerns the \emph{unit of causal attribution}, not simply
the number of turns included in an evaluation.

\paragraph{Search bias, manipulation, and generative echo chambers.} A long
tradition in human-computer interaction and science and technology studies
examines how ranking and presentation shape users'
beliefs~\cite{introna2000shaping,epstein2015search,pariser2011filter}, and
recent work shows LLM-powered search can narrow information
seeking~\cite{sharma2024generative,shah2022situating}. We make the dynamical
mechanism explicit and tie it to a quantitative evaluation gap for GEO.

\section{A Model of Multi-turn GEO}
\label{sec:model}

\subsection{Objects}
Let $S$ be a set of candidate sources. At turn $t = 1,\dots,T$ the system has
two layers.

\emph{Agent layer.} Let $A$ denote the agent operator. The answer $r_t$
assigns each source $s \in S$ a per-turn
visibility $w_t(s) \in [0,1]$, computed by any existing GEO visibility metric
applied turn-by-turn (e.g., position-adjusted word count or an LLM impression
score). We write $w_t(s) = 1$ for a source that monopolizes the turn and
$\sum_s w_t(s) = 1$ when visibility is normalized as a citation share.

\emph{User layer.} The user holds a belief or information-need state $b_t$.
Let $Q$ denote the query operator and $U$ the belief-update operator. The user
generates a query $q_t = Q(b_t)$, reads $r_t$, and updates the state as
$b_{t+1} = U(b_t, r_t)$.

\subsection{Closed-loop dynamics}
The full system is
\begin{equation}
q_t = Q(b_t), \qquad
r_t = A\!\left(q_t,\, h_{t-1},\, \Sgeo\right), \qquad
b_{t+1} = U(b_t, r_t),
\label{eq:loop}
\end{equation}
where $h_{t-1} = (q_1,r_1,\dots,q_{t-1},r_{t-1})$ is the dialogue history and
$\Sgeo$ denotes the set of GEO-optimized sources. GEO enters
the system through the map $S \to \Sgeo$ and biases $A$; our analysis asks how
that initial bias propagates through
Equation~\eqref{eq:loop}. Figure~\ref{fig:overview}(A) depicts the loop.

\subsection{Two independent capture channels}
We decompose capture into two channels that can be measured separately. This
separation is the methodological core of the paper.

\begin{description}
  \item[\textnormal{\textbf{M1 (machine-side capture).}}] Retrieval and
  generation condition on $h_{t-1}$, so a source cited earlier is more likely
  to be cited again \emph{even if the user's query is held fixed}. M1 requires
  no user model: it is a property of $A$ alone and is cleanly measurable by
  comparing a history-conditioned engine to a stateless one on identical
  queries.
  \item[\textnormal{\textbf{M2 (human-side capture).}}] A GEO-shaped answer
  shifts $b_{t+1}$ and therefore $q_{t+1}$ toward the captured source. M2
  operates only through the user's question-asking behavior $Q,U$ and is the
  channel specific to human-agent interaction.
\end{description}

\noindent Separating M1 and M2 supports distinct measurements of the two
mechanisms. M1 is a system-level quantity that does not require a user model,
whereas M2 captures the contribution of human behavior. Each channel can
therefore be evaluated with its own counterfactual
(Section~\ref{sec:illustration}).

\section{Trajectory-Level Evaluation Constructs}
\label{sec:constructs}

We define a set of computable quantities that can be estimated from the
per-turn visibilities $w_t(s)$ produced by an engine along a trajectory.

\subsection{Cumulative conversational visibility (CCV)}
The total visibility a source accrues over a conversation is
\begin{equation}
\CCV_s = \sum_{t=1}^{T} \gamma_t \, w_t(s),
\label{eq:ccv}
\end{equation}
where $\gamma_t \ge 0$ is an attention/salience weight. Setting $\gamma_t=1$
recovers raw cumulative visibility. A decreasing $\gamma_t$ represents a
primacy effect, in which earlier answers receive greater weight, whereas an
increasing $\gamma_t$ represents recency. Empirical evidence about attention
allocation across a session can inform the choice of $\gamma_t$. We use
$\gamma_t = 1$ as a neutral default and report the weighting as a design
parameter.

\subsection{Trajectory gain and its decomposition}
The single-turn gain of a GEO treatment, measured at one isolated turn with the
query held fixed, is
\begin{equation}
L_1(s) = w\!\left(s \mid \text{GEO}\right) - w\!\left(s \mid \neg\text{GEO}\right)
\qquad (\text{single turn, query fixed}),
\label{eq:l1}
\end{equation}
the quantity reported by current single-turn evaluation.
The \emph{trajectory gain}, evaluated along the realized, coupled
conversation, is
\begin{equation}
L_T(s) = \CCV_s(\text{GEO}) - \CCV_s(\neg\text{GEO}).
\label{eq:lt}
\end{equation}
We decompose this gain as
\begin{equation}
L_T \;=\; \underbrace{\Ldir}_{\substack{\text{per-turn lift,}\\\text{query path fixed}}}
\;+\; \underbrace{\Lfb}_{\substack{\text{extra visibility from}\\\text{GEO-shifted queries}}}.
\label{eq:decomp}
\end{equation}
$\Ldir$ is what a single-turn evaluator would (correctly) extrapolate: the
GEO lift summed over turns when the conversational path is held to its
no-GEO counterfactual. $\Lfb$ is the visibility a source gains \emph{because}
GEO changed the subsequent queries and history. Formally, $\Ldir$ and $\Lfb$
are the natural direct and indirect effects of the GEO treatment when the query
and history path acts as a mediator. The decomposition is \emph{path-specific}:
it holds the mediator to its no-GEO realization, and a different reference path
would partition $L_T$ differently. We use the no-GEO path to isolate the
visibility that arises only when the loop is closed. This choice makes $\Lfb$
a well-defined estimand.

\begin{quote}
\textbf{Key property.} $\Lfb \equiv 0$ in any single-turn evaluation and in any
memoryless engine; it is nonzero exactly when a human or a history closes the
loop. $\Lfb$ is the ``human-in-the-loop'' term omitted by current GEO
evaluation.
\end{quote}

\subsection{Separating the two channels: \texorpdfstring{$\Lfb=\LMone+\LMtwo$}{Lfb = LM1 + LM2}}
\label{sec:channelsplit}
The decomposition~\eqref{eq:decomp} isolates the closed loop, but its single
feedback term combines the two channels defined in Section~\ref{sec:model}.
Feedback passes through two mediators in~\eqref{eq:loop}: the dialogue history
$h_{t-1}$ that conditions the agent (M1) and the query path $q_t$ produced by
the user's belief updates (M2). We set these mediators independently in a
counterfactual analysis. Let $w_t^{(h,q)}(s)$ denote per-turn visibility. The
history supplied to $A$ is either its GEO realization ($h{=}1$) or its no-GEO
counterfactual ($h{=}0$). Independently, the user's belief and query path either
drifts under GEO ($q{=}1$) or remains fixed at its no-GEO counterfactual
($q{=}0$). The within-turn GEO bias remains active throughout. Thus,
$w_t^{(0,0)}$ is the direct-only path in~\eqref{eq:decomp}, and
$w_t^{(1,1)}$ is the realized trajectory. We define
\begin{align}
\LMone &= \sum_{t=1}^{T}\gamma_t\bigl[w_t^{(1,0)}(s)-w_t^{(0,0)}(s)\bigr],
\label{eq:lm1}\\
\LMtwo &= \sum_{t=1}^{T}\gamma_t\bigl[w_t^{(1,1)}(s)-w_t^{(1,0)}(s)\bigr].
\label{eq:lm2}
\end{align}
$\LMone$ activates the history channel while holding the user's queries to the
path they would have followed without GEO. It measures the capture produced by
conditioning on the engine's own GEO-influenced history and requires no user
model. $\LMtwo$ then allows the queries to drift while the history remains at
its GEO value. It measures the additional capture produced when a human asks
follow-up questions in a GEO-influenced direction. The endpoints of this
counterfactual chain are the realized and direct-only trajectories, so the
terms telescope exactly:
\begin{equation}
\Lfb=\LMone+\LMtwo.
\label{eq:channelsum}
\end{equation}
M1 can be measured from the system alone by comparing a history-conditioned
engine with a stateless engine on identical queries. M2 is the increment
produced by the human component of the loop and equals the residual
$\Lfb-\LMone$ estimated by the protocol in Section~\ref{sec:limitations}. The
nesting places history before queries so that $\LMone$ is identifiable without
behavioral assumptions. The human-side residual then includes the interaction
between the machine and human channels. The supplementary material
(Section~S6) presents a symmetric attribution of main effects and their
interaction and proves that $\LMone,\LMtwo\ge0$ under confirmatory
reinforcement.

\subsection{Capture coefficient}
The self-reinforcement strength of a source is
\begin{equation}
\kappa(s) = P\!\left(\text{cite}_{t+1}\ni s \mid \text{cite}_t \ni s\right)
          - P\!\left(\text{cite}_{t+1}\ni s \mid \text{cite}_t \not\ni s\right).
\label{eq:kappa}
\end{equation}
$\kappa(s) > 0$ means that a previous citation increases the probability of a
future citation, which is the operational signature of capture. We show below
that GEO raises $\kappa$.

\subsection{Compounding ratio}
The ratio of realized trajectory gain to the naive single-turn extrapolation
is
\begin{equation}
\rho = \frac{L_T}{T \cdot L_1}.
\label{eq:rho}
\end{equation}
$\rho = 1$ means that turns are independent and single-turn evaluation is
exact. $\rho > 1$ means that GEO compounds across the conversation and that a
linear extrapolation from a single turn underestimates its true impact.

\subsection{Misranking diagnostic}
We also compare the \emph{ordering} of GEO methods under the two paradigms.
Given a set of $m$ methods, let $r^{(1)}$
rank them by single-turn gain $L_1$ and $r^{(T)}$ rank them by trajectory gain
$L_T$. We report Kendall's $\tau$ between $r^{(1)}$ and $r^{(T)}$:
\begin{equation}
\tau = \frac{(\#\,\text{concordant}) - (\#\,\text{discordant})}{\binom{m}{2}}.
\label{eq:tau}
\end{equation}
A low or negative $\tau$ indicates that the ``best'' GEO method under
single-turn evaluation differs from the method that captures conversations.
This discrepancy shows that the single-turn paradigm can produce a
decision-relevant ranking error.

\section{Theoretical Analysis: Path Dependence and Superlinear Gain}
\label{sec:theory}

We analyze capture using reinforcement-process
theory~\cite{pemantle2007survey,johnson1977urn}, which also describes
preferential attachment~\cite{barabasi1999scaling} and economic
lock-in~\cite{arthur1989competing}. For clarity, consider two competing sources:
a GEO-optimized source $g$ and an aggregate competitor $c$. We model citations
as draws from a Pólya-type urn. The urn initially contains $G_0$ ``balls'' for
$g$ and $C_0$ for $c$, with the initial composition representing relevance and
salience at turn~1. At each turn, a source is cited with probability
proportional to its current ball count. Its count then increases by a
reinforcement amount $s_g$ or $s_c$. We split the reinforcement asymmetry that
drives capture additively across the two channels:
\begin{equation}
s_g-s_c=\Delta_{M1}+\Delta_{M2},\qquad \Delta_{M1},\Delta_{M2}\ge0,
\label{eq:reinfsplit}
\end{equation}
where $\Delta_{M1}$ is the increment history-conditioning supplies (M1) and
$\Delta_{M2}$ the increment query drift supplies (M2); a stateless engine sets
$\Delta_{M1}=0$ and a non-drifting user sets $\Delta_{M2}=0$. Full statements and
complete proofs are given in the supplementary material (Sections~S1 to S6).

\begin{proposition}[Path dependence / non-ergodicity]
\label{prop:pathdep}
Under symmetric linear reinforcement ($s_g=s_c=s$), the citation share of
source $g$ is a bounded martingale and converges almost surely to a random
limit $W_\infty \sim \mathrm{Beta}(G_0/s,\, C_0/s)$. The limit depends on the
initial composition, not on a deterministic notion of relevance; thus early
turns determine the long-run outcome. GEO, by raising $G_0$ (a turn-1
advantage), shifts the entire limit distribution upward.
\end{proposition}

This result formalizes the persistence of an early advantage. The process is
non-ergodic, and GEO has its greatest leverage in the \emph{early} turns.

\begin{proposition}[Superlinear cumulative gain]
\label{prop:superlinear}
Under asymmetric reinforcement with $s_g > s_c$ (GEO active on both capture
channels), the expected per-turn visibility $w_t(g)$ is strictly increasing in
$t$ throughout the growth regime, and the trajectory gain $L_T$ is strictly
\emph{convex} in $T$ there, meaning that it is superlinear while capture
develops. The
compounding ratio $\rho_T = L_T/(T\,L_1)$ satisfies $\rho_T > 1$ and increases
monotonically in $T$ toward a \emph{finite} ceiling as visibility saturates.
Thus, $L_T$ is asymptotically linear rather than unboundedly superlinear. For
the closed-form recursion in Section~\ref{sec:illustration}, the ceiling is
exactly $1/(2L_1)$. Equivalently, in the mean-field limit, the share
$y_t = w_t(g)$ obeys $\dot y = y(1-y)(s_g - s_c)/N_t > 0$ for $y \in (0,1)$,
where $N_t$ is the total number of balls at turn $t$.
\end{proposition}

\begin{corollary}[The feedback term is the loop]
\label{cor:feedback}
Under confirmatory reinforcement ($s_g \ge s_c$ on the realized path),
$\Lfb \ge 0$, with equality if and only if the reinforcement is symmetric and
the engine is memoryless (no M1) and queries do not drift (no M2). Hence
$\Lfb > 0$ indicates that a human or a history closes the loop in a direction
that favors capture, and its magnitude measures what single-turn evaluation
misses. The corrective regime $s_g < s_c$ instead gives $\Lfb < 0$
(Section~\ref{sec:theory}, ``the sign of the feedback term'').
\end{corollary}

Proposition~\ref{prop:superlinear} also implies $\kappa(g) > 0$: asymmetric
reinforcement is exactly the statement that having been cited raises the next
citation probability. GEO increases $\kappa$ by increasing $s_g$.

\begin{corollary}[Channel decomposition of the feedback term]
\label{cor:channels}
Under the channel-additive reinforcement~\eqref{eq:reinfsplit} (equivalently
$\beta=\beta_{M1}+\beta_{M2}$ in the recursion of
Section~\ref{sec:illustration}), the nested feedback
terms~\eqref{eq:lm1} and~\eqref{eq:lm2} are each nonnegative,
$\LMone,\LMtwo\ge0$, and sum to the feedback term,
$\LMone+\LMtwo=\Lfb$, with $\LMone=0$ iff the engine is memoryless
($\Delta_{M1}=0$) and $\LMtwo=0$ iff queries do not drift ($\Delta_{M2}=0$).
\end{corollary}

This corollary expresses the channel separation in the framework's quantities,
rather than only in its model (Section~\ref{sec:model}). Each capture term is
signed, equals zero under its corresponding null condition, and together the
terms exhaust the feedback effect. The channels also \emph{interact}
super-additively while capture develops and sub-additively as visibility
approaches its ceiling. The supplementary material (Section~S6) identifies this
interaction and presents a Shapley-symmetric alternative to the nested
decomposition.

\paragraph{The sign of the feedback term.} Corollary~\ref{cor:feedback}
assumes \emph{confirmatory} reinforcement, in which the loop favors the
previously cited source ($s_g \ge s_c$). This behavior is predicted by
history-conditioning, model self-consistency, primacy, and
sycophancy~\cite{liu2024lostmiddle,sharma2023sycophancy}. The sign is an
empirical assumption rather than a tautology. If users instead respond to a captured
answer with \emph{corrective} follow-ups (``is this independently
confirmed?'') that steer toward competitors, the human channel M2 is negative
and $\Lfb < 0$, in which case single-turn evaluation overestimates GEO. In
either case, the single-turn estimate is biased, and the trajectory
decomposition identifies that bias. The empirical protocol in
Section~\ref{sec:limitations} is designed to determine which sign dominates in
deployed systems.

\paragraph{From the urn to a computable instance.} The theory and illustration
represent the same phenomenon with different models. We state
Propositions~\ref{prop:pathdep} and~\ref{prop:superlinear} for the stochastic
urn because it yields the non-ergodic random \emph{limit law} in
Proposition~\ref{prop:pathdep}, which a deterministic model cannot produce.
Section~\ref{sec:illustration} uses a deterministic, closed-form recursion from
the same class of capture dynamics: monotone, deviation-amplifying maps with an
absorbing monopoly at full share. The recursion establishes an explicit
correspondence between GEO's two levers. Here, $\delta$ plays the role of the
initial composition $G_0$, and $\beta$ plays the role of the reinforcement
asymmetry $s_g-s_c$. The supplementary material (Section~S4) directly proves
path dependence and superlinearity for this recursion. The illustration is
therefore a second member of the class rather than a discretization of the urn.

\section{Computational Illustration}
\label{sec:illustration}

We instantiate the model from Section~\ref{sec:theory} in a controlled,
deterministic computation. The illustration shows that every construct in
Section~\ref{sec:constructs} is computable, that the predicted phenomena occur,
and that the two evaluation paradigms can rank methods differently. It
demonstrates the constructs rather than providing evidence about deployed
engines. Every reported value follows from Equation~\eqref{eq:recursion} and
the selected inputs. The computation shows that the omitted quantity can affect
method selection, but it does not estimate the size of that quantity in deployed
systems. Section~\ref{sec:limitations} presents an empirical validation protocol
for open conversational question-answering corpora.

\subsection{Setup}
We use the closed-form recursion introduced above. Its parameters correspond to
the urn's two GEO levers: $\delta$ corresponds to $G_0$, and $\beta$ corresponds
to $s_g-s_c$. Let $w_t$ denote the GEO source's per-turn visibility share,
with neutral relevance share $0.5$. The single-turn (direct) GEO lift is
$\delta = L_1$, so
turn~1 visibility is $w_1 = 0.5 + \delta$. The closed-loop reinforcement then
evolves the share by
\begin{equation}
w_{t+1} = w_t + \beta\,(w_t - 0.5)\,(1 - w_t),
\label{eq:recursion}
\end{equation}
where $\beta \ge 0$ is the reinforcement (capture) strength induced by M1 and
M2 jointly; $\beta = 0$ reproduces a memoryless engine with independent turns.
By Equation~\eqref{eq:reinfsplit}, the strength splits across channels as
$\beta=\beta_{M1}+\beta_{M2}$. Running the recursion at $\beta_{M1}$, with only
the history channel active and queries fixed, and at the full $\beta$
instantiates the nested counterfactuals in~\eqref{eq:lm1}
and~\eqref{eq:lm2}. For the worked example, we use the illustrative split
$\beta_{M1}=0.35$ and $\beta_{M2}=0.25$. The qualitative claims do not depend on
the ratio between these components, which must be estimated by the empirical
protocol.
Three reference trajectories are relevant: the \emph{no-GEO} baseline
($w_t = 0.5$), the \emph{direct-only} counterfactual ($w_t = 0.5+\delta$ for
all $t$, with GEO present but the loop held open), and the \emph{full
closed-loop} trajectory of Equation~\eqref{eq:recursion}. We set $\delta=0.12$
and $\beta=0.6$ for the worked example and $\gamma_t=1$, $T=10$.

\subsection{Computed constructs and predicted behavior}
\begin{figure}[t]
  \centering
  \includegraphics[width=\columnwidth]{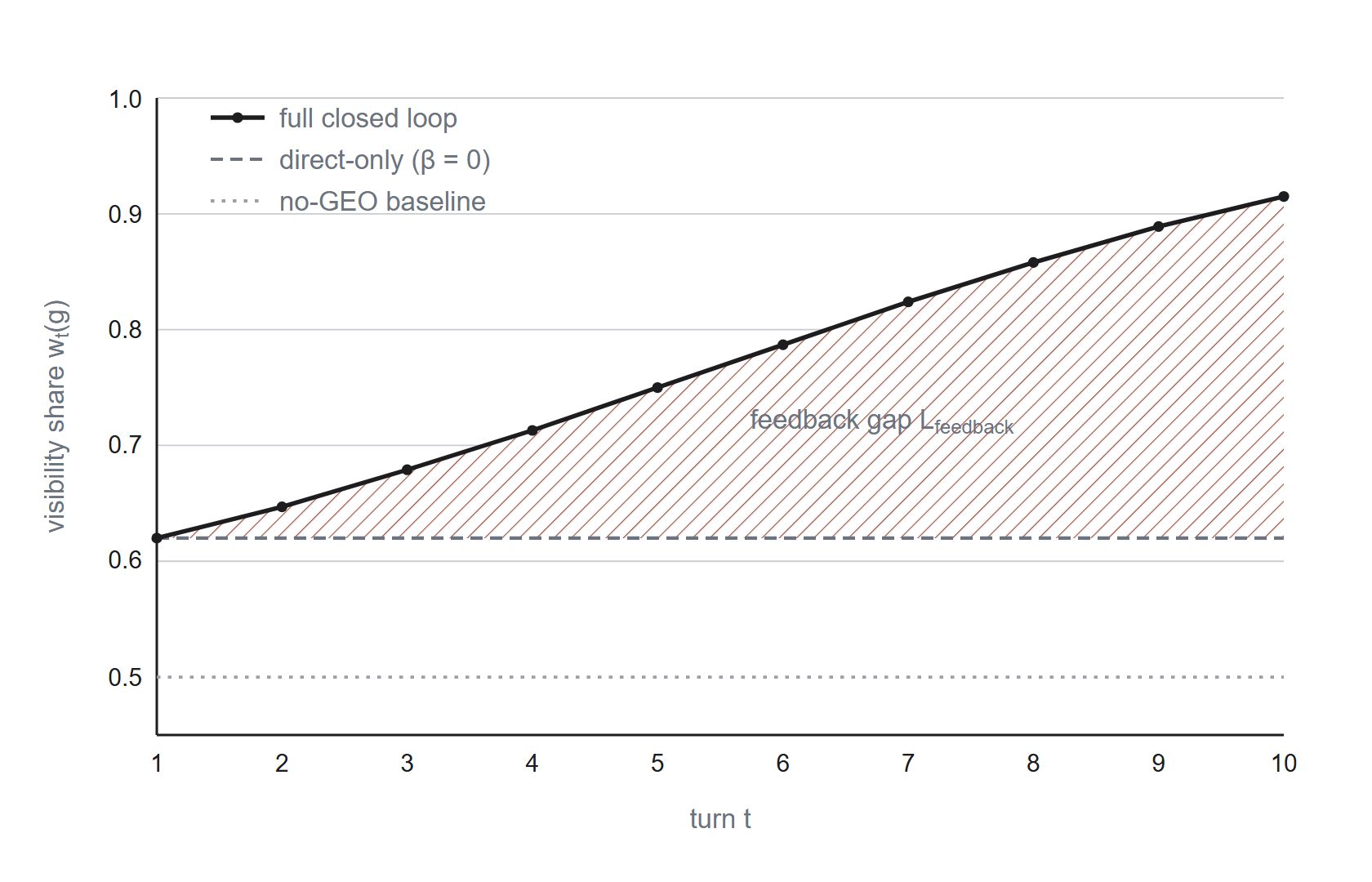}
  \caption{Per-turn visibility of the GEO source under the closed-loop
  recursion~\eqref{eq:recursion} ($\delta=0.12$, $\beta=0.6$). The shaded gap
  between the closed loop and the direct-only counterfactual is the feedback
  contribution $\Lfb$.}
  \Description{A line chart with turn on the x-axis and visibility share on
  the y-axis. The full closed-loop curve rises from 0.62 to 0.91; the
  direct-only line is flat at 0.62; the baseline is flat at 0.5.}
  \label{fig:traj}
\end{figure}

Figure~\ref{fig:traj} plots the three trajectories. The closed-loop share
climbs from $0.62$ at turn~1 to $0.91$ at turn~10, while the direct-only
counterfactual stays flat at $0.62$. The gap is pure feedback, namely the
visibility that GEO produces only because the loop is closed.

Table~\ref{tab:decomp} reports the decomposition at $T=10$. The trajectory
gain $L_T = 2.68$ splits into a direct term $\Ldir = 1.20$ (exactly the
single-turn extrapolation $T\cdot L_1$) and a feedback term $\Lfb = 1.48$. The
feedback term is \emph{larger than} the direct term: more than half of GEO's
modeled conversational payoff is not represented by single-turn evaluation.
Corollary~\ref{cor:channels} further splits the feedback term into a
machine-side component $\LMone=0.85$ and a human-side component $\LMtwo=0.63$.
The machine-side channel, which requires no user model, contributes a feedback
gain ($0.85$) comparable to the direct term ($1.20$). The human component adds
another $0.63$. The compounding ratio is $\rho = 2.23$, indicating more than
twice the gain predicted by a linear extrapolation from a single-turn
measurement. As $T$ grows,
$\rho_T$ continues to climb toward its ceiling $1/(2\delta)\approx 4.17$ (the
saturated per-turn gain $0.5$ divided by the single-turn gain
$\delta$): the underestimation increases monotonically with conversation length
before leveling off, consistent with Proposition~\ref{prop:superlinear}.

\begin{table}[t]
  \caption{Decomposition of trajectory gain at $T=10$
  ($\delta=0.12$, $\beta=0.6$, $\gamma_t=1$). The feedback term splits into its
  machine-side and human-side channels under the illustrative
  $\beta_{M1}=0.35$, $\beta_{M2}=0.25$; the split is exact
  ($0.851+0.630=1.481$).}
  \label{tab:decomp}
  \begin{tabular}{lr}
    \toprule
    Quantity & Value \\
    \midrule
    Single-turn gain $L_1$ & $0.120$ \\
    Naive extrapolation $T\cdot L_1$ & $1.200$ \\
    Direct term $\Ldir$ & $1.200$ \\
    Feedback term $\Lfb$ & $1.481$ \\
    \quad machine-side $\LMone$ ($\beta_{M1}=0.35$) & $0.851$ \\
    \quad human-side $\LMtwo$ ($\beta_{M2}=0.25$) & $0.630$ \\
    Trajectory gain $L_T$ & $2.681$ \\
    Compounding ratio $\rho$ & $2.234$ \\
    \bottomrule
  \end{tabular}
\end{table}

Figure~\ref{fig:overview}(B) confirms Proposition~\ref{prop:superlinear}: the
trajectory gain $L_T$ (solid) bows above the linear single-turn extrapolation
$T\cdot L_1$ (dashed), and the compounding ratio $\rho_T$ rises monotonically
from $1.00$ at $T=1$ to $2.23$ at $T=10$. Single-turn evaluation is exact only
at $T=1$, and its error increases with conversation length.

\subsection{Misranking under single-turn evaluation}
\label{sec:misrank}
We next test whether single-turn and trajectory evaluation \emph{order GEO
methods differently}. We consider five stylized GEO methods that trade off
single-turn salience $\delta$ against induced capture strength $\beta$. This
tradeoff represents the distinction between tactics that maximize immediate
prominence, such as keyword stuffing, and tactics that steer follow-up
questions, such as framing that seeds the user's next query. For each method,
we compute
$L_1=\delta$ and $L_T$ from Equation~\eqref{eq:recursion} at $T=10$.
Table~\ref{tab:misrank} reports the result.

\begin{table}[t]
  \caption{Five GEO methods ranked by single-turn vs.\ trajectory gain
  ($T=10$). Single-turn and trajectory orderings disagree (Kendall's
  $\tau=0.4$); the single-turn winner (A) is only third on trajectory, while
  the trajectory winner (B) is second on single-turn.}
  \label{tab:misrank}
  \begin{tabular}{cccccc}
    \toprule
    Method & $\delta$ ($=L_1$) & $\beta$ & $L_T$ &
      rank$_{L_1}$ & rank$_{L_T}$ \\
    \midrule
    A & $0.18$ & $0.2$ & $2.343$ & 1 & 3 \\
    B & $0.15$ & $0.7$ & $3.224$ & 2 & \textbf{1} \\
    C & $0.12$ & $0.3$ & $1.919$ & 3 & 4 \\
    D & $0.10$ & $0.8$ & $2.851$ & 4 & 2 \\
    E & $0.08$ & $0.5$ & $1.855$ & 5 & 5 \\
    \bottomrule
  \end{tabular}
\end{table}

The two rankings agree on only $7$ of $10$ pairs, giving Kendall's
$\tau = (7-3)/10 = 0.4$, which indicates at most a weak correlation. Method~A
ranks first under single-turn evaluation but only third by trajectory impact.
Method~B ranks second on a single turn but first by trajectory impact, while
method~D ranks next to last by single-turn gain but second by trajectory
impact. A practitioner who optimizes the single-turn metric would therefore
select the wrong method within this model. Whether deployed GEO methods are
sufficiently separated in the $(\delta,\beta)$ plane to produce such
misranking remains an empirical question (Section~\ref{sec:limitations}).

\subsection{Robustness of the qualitative claims}
The specific numbers depend on $\delta$, $\beta$, and $T$, but the qualitative
claims do not. For any $\beta>0$, Equation~\eqref{eq:recursion} implies
$\Lfb>0$ and $\rho>1$ because $w_t$ strictly increases while the direct-only
baseline remains flat. Whenever single-turn salience and capture strength are
not perfectly aligned across methods, the single-turn and trajectory rankings
can diverge ($\tau<1$).

\paragraph{Misranking across random method sets.} To assess whether
Table~\ref{tab:misrank} depends on the five selected methods, we draw
$5\times10^4$ random sets of five methods, with $\delta\sim\mathcal{U}[0.04,0.24]$
and $\beta\sim\mathcal{U}[0.05,1.0]$ drawn \emph{independently}, and recompute
the single-turn/trajectory Kendall $\tau$ at $T=10$. The rankings disagree
($\tau<1$) in $90\%$ of draws (mean $\tau=0.54$, median $0.60$), and in $48\%$
the single-turn \emph{winner} is not the trajectory winner. When $\beta$ is
redrawn to be negatively correlated with $\delta$, representing the proposed
salience and capture tradeoff, the disagreement increases to a mean
$\tau=0.18$, with the single-turn winner wrong in $91\%$ of draws. The
worked example is therefore representative of the model's behavior under these
sampling assumptions. Empirical evaluation must estimate where
deployed GEO methods sit in the $(\delta,\beta)$ plane
(Section~\ref{sec:limitations}).

\section{Human-Computer Interaction Foundations}
\label{sec:hci}

The human-side channel M2 makes the query endogenous and closes the loop. We
ground each user-layer operator in an established theory so that the human
component corresponds to a specified mechanism.

\paragraph{The query operator $Q$ via information foraging.}
Information foraging theory models the user as a forager who follows a gradient
of \emph{information scent}, the perceived cues about where value
lies~\cite{pirolli1999infoforaging}. The follow-up query
$q_{t+1}=Q(b_{t+1})$ is the forager's next move along that gradient. A
GEO-shaped answer can increase the perceived value of one source's region and
direct the foraging path toward it. This interpretation gives $Q$ a concrete,
testable form and explains why queries drift toward captured sources.

\paragraph{The consequence layer via trust calibration.}
Appropriate reliance requires perceived trustworthiness to track actual
reliability~\cite{lee2004trust}. Conversational capture separates the two.
Repeated exposure can increase a source's perceived reliability through
familiarity while reducing exposure to alternative sources that could support
recalibration. Capture can therefore \emph{amplify} trust miscalibration. At the
model layer, miscalibration is the gap between a source's citation share and its
ground-truth reliability share. Measuring this construct with real users
remains future work.

\paragraph{The normative layer via Bayesian persuasion.}
Multi-turn GEO is a sequential persuasion game~\cite{kamenica2011bayesian}:
the content owner is a sender choosing a signaling scheme (the GEO treatment),
the agent is the channel, and the user is a receiver updating a posterior.
This framing provides the normative concepts of welfare, manipulation, and
commitment without assuming that users are manipulated. It treats GEO as a
designed information-disclosure strategy whose multi-turn form is strictly more
powerful than its single-turn form.

Together, these three established theories specify the roles of the human
components in Equation~\eqref{eq:loop} for the multi-turn GEO setting.

\section{Discussion: Implications for Answer-Engine Design}
\label{sec:discussion}

\paragraph{Conversational capture as a manipulation surface.} Single-turn GEO
competes for one answer, whereas multi-turn GEO competes for a trajectory.
Proposition~\ref{prop:pathdep} shows that its leverage is concentrated in the
\emph{earliest} turns. Defenses should therefore focus on those turns.

\paragraph{Design levers.} The model suggests three mitigations, each of which
targets a term in the framework. First, \emph{per-turn re-grounding} reduces the
weight of $h_{t-1}$ during retrieval, thereby addressing M1 and lowering
$\kappa$. Second, \emph{enforced source diversity} constrains the share of any
single source within a turn and directly bounds $\Lfb$. Third, \emph{provenance
transparency} explains why a source continues to be cited, distinguishing a
previous citation from current relevance. This information gives the user a
corrected information scent and supports trust recalibration.

\paragraph{Evaluation should move to trajectories.} GEO
audits, answer-engine benchmarks, and platform-side abuse detection should
report trajectory-level quantities ($L_T$, $\Lfb$, $\rho$, $\kappa$) and the
single-turn/trajectory misranking $\tau$, not single-turn visibility alone.
These constructs use the same per-turn $w_t(s)$ values produced by existing
pipelines.

\section{Limitations and Threats to Validity}
\label{sec:limitations}

We discuss three threats to validity and describe an empirical protocol for
evaluating them.

\paragraph{Computability and empirical validation.} Every construct in
Section~\ref{sec:constructs} is computable from per-turn visibilities, and the
misranking result (Section~\ref{sec:misrank}) gives the single-turn paradigm's
decision relevance a concrete failure mode. The next empirical step is to
estimate $w_t(s)$, $\kappa$, $\rho$, and $\tau$ on conversational
open-retrieval question-answering corpora that provide human follow-up
sequences and permit corpus modification. Suitable corpora include
TopiOCQA~\cite{adlakha2022topiocqa}, QReCC~\cite{anantha2021qrecc}, and
ORConvQA~\cite{qu2020orconvqa}. The protocol would inject GEO-optimized and
control passages and run an open-source RAG pipeline turn by turn along the
observed query sequences.

\paragraph{Dependence on user simulation.} In the
planned empirical study, the \emph{main trajectory} is anchored by real human
queries, and the machine-side channel M1 is measured \emph{without any user
model}. Comparing a history-conditioned engine with a stateless engine on
identical queries directly estimates $\LMone$ in
Equation~\eqref{eq:lm1}. The human-side $\LMtwo$ is then the residual
$\Lfb-\LMone$, recovered from the M2 counterfactual (``what the user would
have asked absent GEO bias''). Only this counterfactual requires a simulated
branch. That branch should use a calibrated LLM user simulator whose follow-up
distribution matches a taxonomy of observed follow-up moves, including
deepening, rephrasing, challenging, and pivoting, mined from interaction
logs~\cite{zhao2024wildchat,zheng2024lmsys}. The analysis should report
sensitivity to the simulator. The values in this paper are explicitly derived
from the model and do not replace this study. Because M1 requires neither a user
model nor a simulated branch, it can already be measured on these corpora by
comparing a history-conditioned engine with a stateless engine on the observed
query sequences. This measurement would replace the illustrative
$\beta_{M1}/\beta_{M2}$ split with an empirical estimate and is the immediate
next step.

\paragraph{Behavioral assumptions.} $Q$, $U$, and the trust-miscalibration
construct are each grounded in an established
theory~\cite{pirolli1999infoforaging,lee2004trust,kamenica2011bayesian}. We do
\emph{not} claim that users are manipulated. The model instead yields a
hypothesis that requires behavioral validation.

\paragraph{Other limitations.} (i) LLM-as-judge visibility scoring is
imperfect; an empirical study should report agreement with gold labels and
inter-judge reliability~\cite{zheng2023judging}. (ii) Open-source RAG is not a
commercial closed engine; replicating across several open models guards
against single-model artifacts and is itself a reproducibility asset. (iii) We
measure no real human trust or perception; that is future work. (iv) The
two-source urn collapses a many-source landscape into one aggregate
competitor. Under linear reinforcement the martingale argument and the
non-ergodic limit generalize to $K$ sources, with the Beta limit becoming a
Dirichlet distribution~\cite{pemantle2007survey}. This result appears in the
supplementary material (Section~S2), but competition \emph{among} several
optimized sources is
outside the model.
\section{Conclusion}
\label{sec:conclusion}

GEO is currently evaluated as if a conversation were a sequence of independent
search queries, although an answer engine influences the user's next question.
An initial GEO advantage can therefore propagate through a human-agent feedback
loop and compound. We call this phenomenon conversational capture. We
formalized the loop, separated its machine-side and human-side channels, and
defined trajectory-level evaluation constructs. The feedback term $\Lfb$ is
identically zero under single-turn
evaluation. Reinforcement-process theory shows the dynamics are non-ergodic
(early turns decide the outcome) and that cumulative GEO gain grows
superlinearly with conversation length while capture develops. Single-turn
extrapolation therefore underestimates the gain by a margin that widens with the
conversation. A model-derived computation shows that the feedback term can
dominate the direct term and that single-turn and trajectory rankings of GEO
methods can diverge. Single-turn evaluation can consequently select the wrong
method. Evaluation should instead measure trajectories, while answer engines
should re-ground, diversify, and explain persistent citations. The next step is
to validate the human-side channel empirically with conversational corpora and
real users.

\begin{acks}
This research was supported by JSPS KAKENHI Grant Number 25K21201.
\end{acks}

\bibliographystyle{ACM-Reference-Format}
\bibliography{references}

\end{document}